\documentclass[pre,aps,12pt]{revtex4}

\usepackage{graphicx}
\usepackage{color}

\begin{document}

\title{Delta chain with ferromagnetic and antiferromagnetic interactions
at the critical point}
\author{V.Ya.\ Krivnov, D.V.\ Dmitriev}
\affiliation{Joint Institute of Chemical Physics of RAS, Kosygin
str.4, 119991, Moscow, Russia.}

\author{S.\ Nishimoto, S.-L.\ Drechsler}
\affiliation{Leibniz-Institut fuer Festk\"orper- und
Werkstoffforschung Dresden, D-01171 Dresden, Germany}
\author{J.\ Richter}
\affiliation{Institut f\"ur Theoretische Physik, Universit\"at
Magdeburg, P.O. Box 4120, D-39016, Magdeburg, Germany}
\date{}

\begin{abstract}
We investigate the spin-1/2 Heisenberg model on the delta chain
(sawtooth chain) with ferromagnetic nearest-neighbor  and
antiferromagnetic next-neighbor  interactions. For a special ratio
between these interactions there is a class of exact ground states
formed by localized magnons and the ground state is
macroscopically degenerate with a large residual entropy per spin
$s_0=\frac{1}{2}\ln 2$. An important feature of this model is a
sharp decrease of the gaps for excited states with an increase of
the number of magnons. These excitations give an essential
contribution to the low-temperature thermodynamics. The behavior
of the considered model is compared with that of the delta chain
with both antiferromagnetic interactions.
\end{abstract}

\maketitle

\section{Introduction}
Quantum many-body systems with a single-particle flat band have
attracted much attention. About twenty years ago Mielke and Tasaki
\cite{mielke,tasakiPRL,tasakiCMP,maksym2012} showed that a
repulsive on-site interaction in flat-band Hubbard systems  yields
ferromagnetic ground states. More recently, a very active and
still ongoing discussion of flat-band systems in the context of
topological insulators has been started, see, e.g.
Ref.~\onlinecite{bergholtz} and references therein. Frustrated
quantum antiferromagnets represent another active research field,
where flat-band physics my lead to interesting low-temperature
phenomena
\cite{schnack,Schulenburg,Richter2004,Zhitomir,derzhko,derzhko2006,tsunetsugu},
such as a macroscopic jump in the ground-state magnetization curve
and a nonzero residual ground-state entropy at the saturation
field as well as an extra low-temperature peak in the specific
heat. All these phenomena are related to the existence of a class
of exact eigenstates in a form of localized multi-magnon states
which become ground states in high magnetic fields.

An interesting and typical example of such a flat-band system is
the $s=\frac{1}{2}$ delta or sawtooth Heisenberg model consisting
of a linear chain of triangles as shown in Fig.\ 1. The
interaction $J_{1}$ acts between the apical (even) and the basal
(odd) spins, while $J_{2}$ is the interaction between the neighbor
basal sites. There is no direct exchange between apical spins. The
Hamiltonian of this model has the form
\begin{equation}
\hat{H}=J_{1}\sum (\mathbf{S}_{2n-1}\cdot
\mathbf{S}_{2n}+\mathbf{S}_{2n}\cdot
\mathbf{S}_{2n+1}-\frac{1}{2})+J_{2}\sum (\mathbf{S}_{2n-1}\cdot
\mathbf{S}_{2n+1}-\frac{1}{4})-h\sum S_{n}^{z} , \label{q}
\end{equation}
where $\mathbf{S}_{n}$ are $s=\frac{1}{2}$ operators and $h$ is
the dimensionless magnetic field.

The ground state of model (\ref{q}) with both antiferromagnetic
$J_{1}>0$ and $J_{2}>0$ (AF delta chain) has been studied as a
function of $J_2/J_1$ in Refs.\onlinecite{nakamura,sen,blundell}.
At high magnetic fields  for excitations above the fully polarized
ferromagnetic state the lower one-magnon band  is dispersionless
for a special choice of the coupling constants $J_2=J_1/2$
\cite{derz}. The excitations in this band are localized states,
i.e. the excitations are restricted to a finite region of the
chain. These localized one-magnon states allow to construct a set
of multi-magnon states. Configurations, where the localized
magnons spatially separated (isolated) from each other, become
also exact eigenstates of the Hamiltonian (\ref{q}). At the
saturation field $h=h_{s}=2J_{1}$ all these states have the lowest
energy and the ground state is highly degenerated
\cite{derzhko,derz,Zhitomir}. The degree of the degeneracy can be
calculated by taking into account a hard-core rule forbidding the
overlap of localized magnons with each other (hard-dimer rule).
Exact diagonalization studies\cite{derzhko2006,derz} indicate,
that the ground states in this antiferromagnetic model are
separated by finite gaps from the higher-energy states. Thus the
localized multi-magnon states can dominate the low-temperature
thermodynamics in the vicinity of the saturation field and the
thermodynamic properties can be calculated by mapping the AF delta
chain onto the hard-dimer problem \cite{Zhitomir,derzhko,derz}. A
similar structure of the ground states with localized magnons is
realized in a variety of frustrated spin lattices in one, two and
three dimensions such as the kagome, the checkerboard, the
pyrochlore lattices, see e.g.
Refs.\onlinecite{Schulenburg,Richter2004,Zhitomir,derzhko,derzhko2006,tsunetsugu}.

In contrast to the AF delta chain, the model (\ref{q}) with
ferromagnetic $J_{1}<0$ and antiferromagnetic $J_{2}>0$
interactions ( F-AF delta chain) is less studied, though it is
rather interesting. In particular, it is a minimal model for the
description of the quasi-one-dimensional compound
$[Cu(bpy)H_{2}O][Cu(bpy)(mal)H_{2}O](ClO_{4})$ containing magnetic
$Cu^{2+}$ ions  \cite{Inagaki}.

It is known \cite{Tonegawa} that the ground state of the F-AF
delta chain is ferromagnetic for $\alpha =\frac{J_{2}}{\left\vert
J_{1}\right\vert }<\frac{1}{2}$. In Ref.~\onlinecite{Tonegawa} it
was argued that the ground state for $\alpha>\frac{1}{2}$ is a
special ferrimagnetic state. The critical point $\alpha
=\frac{1}{2}$ is the transition point between these two ground
state phases.

In this paper we will demonstrate that the behavior of the model
at this point is highly non-trivial. Similarly to the AF delta
chain also the F-AF model at the critical point supports localized
magnons which are exact eigenstates of the Hamiltonian. They are
trapped in a valley between two neighboring triangles, where the
occupation of neighboring valleys is forbidden (the so-called
non-overlapping or isolated localized-magnon states.) We will show
that the ground states in the spin sector $S=S_{\max}-k$, $k<N/4$,
consist of states with $k$ isolated localized magnons ($k$-magnon
states), but in contrast to the AF case they are exact ground
states at zero magnetic field \cite{remark}. Moreover, in addition
to $k$-magnon configurations consisting of non-overlapping
localized magnons there are states with overlapping ones. Hence,
the degree of degeneracy of the ground state is even larger than
in the AF delta chain. Another difference to the localized-magnon
states in the AF delta chain concerns the gaps between the ground
state and the excited states which become very small for $k>1$. It
means that the contribution of the ground states to the
thermodynamics does not dominate even for low temperatures.

Our paper is organized as follows. In Section II we consider the
ground states of the F-AF delta chain at the critical point. Based
on the localized-states scenario we calculate analytically the
degree of the ground-state degeneracy and check our analytical
predictions by comparing them with full exact diagonalization (ED)
data for finite chains up to $N=24$ sites. In the Section III we
study the low-temperature thermodynamics of the considered model.
We will show that the low-lying states are separated from the
ground states by very small gaps. These low-lying excitations give
the dominant contribution to the thermodynamics as the temperature
grows from zero and approaches these small gaps. We calculate
different thermodynamic quantities, such as magnetization,
susceptibility, entropy, and specific heat by full ED of finite
chains and discuss the low-temperature behavior of these
quantities. In Section IV we consider the magnetocaloric effect in
the critical F-AF delta chain. In the concluding section we give a
summary of our results.

\begin{figure}[tbp]
\includegraphics[width=5in,angle=0]{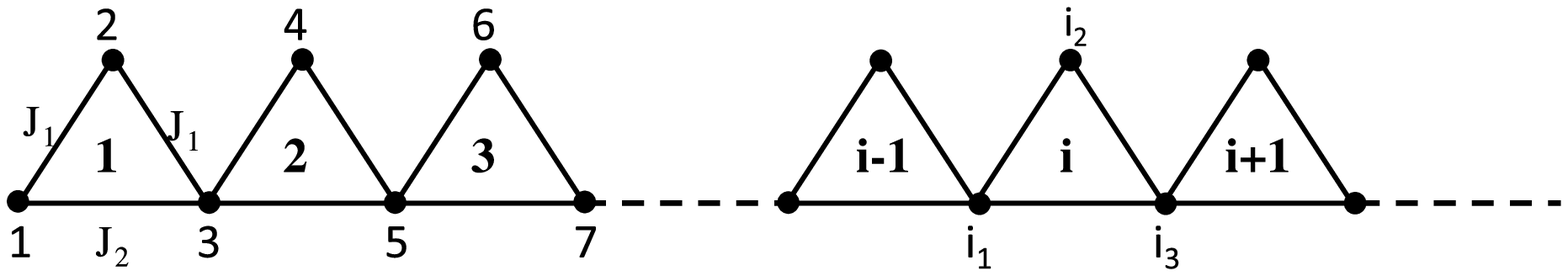}
\caption{The $\triangle$-chain model.} \label{fig1}
\end{figure}

\section{Ground state}

In this section we study the ground state of the F-AF delta chain
at the critical point. For this aim it is convenient to represent
the Hamiltonian (\ref{q}) at $\alpha =\frac{1}{2}$ as a sum of
local Hamiltonians
\begin{equation}
\hat{H}=\sum \hat{H}_{i}  \label{q1}
\end{equation}%
where $\hat{H}_{i}$ is the Hamiltonian of the $i$-th triangle,
which can be written in a form
\begin{equation}
\hat{H}_{i}=-(\mathbf{S}_{i_1}+\mathbf{S}_{i_3})\cdot
\mathbf{S}_{i_2}\mathbf{+}\frac{1}{2}\mathbf{S}_{i_1}\cdot
\mathbf{S}_{i_3}+\frac{3}{8} .\label{q2}
\end{equation}
In Eq.(\ref{q2}) we put $J_{1}=-1$.
The three eigenvalues of Eq.(\ref{q2}) are $E_{i}=0$, $E_{i}=0$
and $E_{i}=\frac{3}{2}$ for the states with spin quantum numbers $S=\frac{3}{2}$,
$S=\frac{1}{2}$ and $S=\frac{1}{2}$, correspondingly. Because the
local Hamiltonians $\hat{H}_{i}$ generally do not commute with
each other, for the lowest eigenvalue $E_{0}$ of $\hat{H}$ holds
\begin{equation}
E_{0}\geq \sum E_{i}=0 .  \label{q3}
\end{equation}
It is evident that the energy of the ferromagnetic state with
maximal total spin $S_{\max}=\frac{N}{2}$ of model
(\ref{q1}) is
zero. Therefore, the inequality in Eq.(\ref{q3}) turns in an
equality and the ground state energy of Eq.~(\ref{q1}) is zero. The
question is: how many states with different total spin have zero
energy?

At first, we consider one-magnon states with $S=S_{\max }-1$. The
spectrum $E(q)$ of these states for the F-AF delta chain with
periodic boundary conditions (PBC) has two branches. One of them
is dispersionless with $E(q)=0$ while the second branch is
dispersive and its energy is
\begin{equation}
E(q)=2-\sin ^{2}q,\quad -\frac{\pi }{2}<q<\frac{\pi }{2} . \label{one-magnon}
\end{equation}
The dispersionless one-magnon states correspond to localized
states which can be chosen as
\begin{equation}
\hat{\varphi} _{1}\left\vert F\right\rangle
=(s_{2}^{-}+s_{4}^{-}+2s_{3}^{-})\left\vert F\right\rangle ,\; \hat{\varphi}
_{2}\left\vert F\right\rangle =(s_{4}^{-}+s_{6}^{-}+2s_{5}^{-})\left\vert
F\right\rangle ,\; \ldots \; , \hat{\varphi} _{n}\left\vert F\right\rangle
=(s_{N}^{-}+s_{2}^{-}+2s_{1}^{-})\left\vert F\right\rangle  \label{q4}
\end{equation}%
where $n=\frac{N}{2}$ and $\left\vert F\right\rangle =\left\vert \uparrow
\uparrow \uparrow \ldots \uparrow \right\rangle $.
These functions are exact eigenfunctions of each local
$\hat{H}_{i}$ with zero energy. It can be checked directly that
$\hat{H}_{l}\hat{\varphi} _{l}\left\vert F\right\rangle =0$ and
$\hat{H}_{l+1}\hat{\varphi} _{l}\left\vert F\right\rangle =0$,
while for other $i\neq l-1,l$ the local Hamiltonian $\hat{H}_{i}$
and the operators $\hat{\varphi} _{l}$ defined by Eq.(\ref{q4})
commute giving $\hat{H}_{i}\hat{\varphi} _{l}\left\vert
F\right\rangle =\hat{\varphi} _{l}\hat{H}_{i}\left\vert
F\right\rangle =0$. The $n$ states (\ref{q4}) form a complete
nonorthogonal basis in the space of the dispersionless branch. It
follows from the fact that the relation
\begin{equation}
\sum a_{i}\hat{\varphi} _{i}=0
\end{equation}%
is fulfilled if all $a_{i}=0$, only. Besides, we note that there
are $(n-1)$ linear combinations of $\hat{\varphi} _{i}\left\vert
F\right\rangle $ which belong to the states with $S=S_{\max }-1$
and one combination belongs to $S=S_{\max }$. The latter is
\begin{equation}
\sum \hat{\varphi} _{i}\left\vert F\right\rangle =2S_{tot}^{-}\left\vert
F\right\rangle .
\end{equation}

For the F-AF delta chain with open boundary conditions (OBC)
and odd $N$ there are $n=\frac{N+1}{2}$ localized one-magnon
states with zero energy and their wave functions are
\begin{equation}
\hat{\varphi} _{1}\left\vert F\right\rangle
=(s_{2}^{-}+2s_{1}^{-})\left\vert F\right\rangle ,\; \hat{\varphi}
_{2}\left\vert F\right\rangle
=(s_{2}^{-}+s_{4}^{-}+2s_{3}^{-})\left\vert F\right\rangle , \;
\ldots \; , \hat{\varphi} _{n}\left\vert F\right\rangle
=(s_{N-1}^{-}+2s_{N}^{-})\left\vert F\right\rangle . \label{q5}
\end{equation}
These functions are linearly independent similarly to those for
the periodic delta chain. It is convenient to introduce another
set of linearly independent operator functions instead of
$\hat{\varphi} _{i}$ which have the form
\begin{equation}
\hat{\Phi} (m)=\sum_{i=1}^{m}\hat{\varphi} _{i},\quad m=1,2\ldots n
\end{equation}
All functions $\hat{\Phi} (m)\left\vert F\right\rangle $ are
eigenfunctions with zero energy of each local Hamiltonian
$\hat{H}_{i}$. Similarly to the periodic chain the $(n-1)$
functions $\hat{\Phi} (m)\left\vert F\right\rangle $ with
$m=1,2,..,n-1$ belong to $S=S_{\max }-1$ and $\hat{\Phi}
(n)\left\vert F\right\rangle $ is the function of the state with
$S=S_{\max }$ and $S^{z}=S_{\max }-1$ because $\hat{\Phi}
(n)=2S_{tot}^{-}$.

Let us consider two-magnon states. For simplicity we will deal
with the delta chain with OBC. It is clear that the pair of
isolated (non-overlapping) magnons is an exact ground state of
the Hamiltonian {(\ref{q1}) and the wave functions of pairs, $\hat{\varphi}
_{i}\hat{\varphi}_{j}\left\vert F\right\rangle $ $(j\geq i+1)$ are
exact ground state functions of each local $\hat{H}_{l}$ with zero
energy. The number of such pairs is $C_{n-1}^{2}$, where
 $C_{m}^{n}=\frac{m!}{n!(m-n)!} $ is the binomial coefficient. It
can be proved similarly to the case of the AF delta chain
\cite{Richter} that these states are linearly independent.

In fact, the exact two-magnon ground state wave functions of the
Hamiltonian (\ref{q1}) at $\alpha =\frac{1}{2}$ can be chosen by
many other ways. We determine the set of two-magnon states as
following
\begin{equation}
\hat{\Phi} (m_{1})\hat{\Phi} (m_{2})\left\vert F\right\rangle ,\quad 1\leq
m_{1}<m_{2}\leq n-1 . \label{q6}
\end{equation}
Though Eq.\ (\ref{q6}) contains products of interpenetrating
operator functions $\hat{\varphi} _{i}$ (i.e.\ acting on commonly
involved sites), it is easy to be convinced that the states
defined in  Eq.\ (\ref{q6}) are exact ground state wave functions
of each $\hat{H}_{l}$. For example, let us consider the function
$\hat{\Phi} (1)\hat{\Phi} (2)\left\vert F\right\rangle $. It
equals
\begin{equation}
\hat{\Phi} (1)\hat{\Phi} (2)\left\vert F\right\rangle =(\hat{\varphi} _{1}+\hat{\varphi}
_{2})\hat{\varphi} _{1}\left\vert F\right\rangle
=(2s_{1}^{-}+2s_{2}^{-}+2s_{3}^{-}+s_{4}^{-})\hat{\varphi}
_{1}\left\vert F\right\rangle =(2S^{-}(1)+s_{4}^{-})\hat{\varphi}
_{1}\left\vert F\right\rangle , \label{q7}
\end{equation}%
where $S^{-}(1)$ is the lowering spin operator of the first
triangle. Then, this function is an exact ground state function of
$\hat{H}_{1}$, because $\hat{\varphi} _{1}$ creates a mixture of
the states with $S=\frac{3}{2}$ and $S=\frac{1}{2}$ of
$\hat{H}_{1}$ with zero energy. On the other hand, this function
is an exact ground state function of $\hat{H}_{2}$, because it
contains the combination $2s_{3}^{-}+s_{4}^{-}$ in the first
bracket. It is also clear that the function (\ref{q7}) is an exact
ground state function of $\hat{H}_{i}$ with $i\geq 3$ because
$\hat{H}_{i}$ for these $i$ commute with $\hat{\Phi} (1)\hat{\Phi}
(2)$ and $\hat{H}_{i}\hat{\Phi}
 (1)\hat{\Phi}
(2)\left\vert F\right\rangle =\hat{\Phi} (1)\hat{\Phi} (2)\hat{H}
_{i}\left\vert F\right\rangle =0$. A similar consideration can be
extended to any function having the form (\ref{q6}). The function
$\hat{\Phi} (m_{1})\hat{\Phi} (m_{2})\left\vert F\right\rangle $
contains the lowering operators $S^{-}(1,2\ldots m_{1}-1)$ and
$S^{-}(1,2\ldots m_{2}-1)$ (where $S^{-}(1,2\ldots k) $ is the
total lowering spin operator for the first $k$ triangles). The
construction of the brackets in Eq.\ (\ref{q6}) ensures the
relation $\hat{H}_{i}\hat{\Phi} (m_{1})\hat{\Phi}
(m_{2})\left\vert F\right\rangle =0$ for $i\leq m_{2}$, while this
relation for $i>m_{2}$ is fulfilled automatically. It easy to
check that the set of functions (\ref{q6}) can be transformed to
the set $\hat{\varphi} _{i}\hat{\varphi} _{j}\left\vert
F\right\rangle $ $(j\geq i+1)$ using the condition $\hat{\Phi}
(n)=2S_{tot}^{-}$.

Strictly speaking we should also show that the set of the states
(\ref{q6}) after a projection onto the states with
$S_{tot}=S^{z}=S_{\max }-2$ gives all linearly independent states
in this spin sector. We checked this analytically for systems with
$n=5,7$ (i.e.\ $N=11,15$) but we did not succeed with a rigorous
proof of this statement.

Since the operator function $\hat{\Phi} (n)$ with $m_{2}\leq n-1$
belongs to a state $\hat{\Phi} (m_{1})\hat{\Phi} (n)\left\vert
F\right\rangle =2S_{tot}^{-}\hat{\Phi} (m_{1})\left\vert
F\right\rangle $ in the sector $S_{tot}=S_{\max }-1$, it is not
described by Eq.\ (\ref{q6}) by definition. The number of states
described by Eq.\ (\ref{q6}) amounts $C_{n-1}^{2}$.

Now we consider the general case of the $k$-magnon subspace with
$S_{tot}=S^{z}=S_{\max }-k$. It is evident that a state consisting
of $k$ isolated localized magnons
\begin{equation}
\hat{\varphi} _{i_1}\hat{\varphi} _{i_2}\hat{\varphi} _{i_3}\ldots \hat{\varphi}
_{i_k}\left\vert F\right\rangle ,\quad i_{l}>i_{l-1}+1
  \label{k0-set}
\end{equation}%
is an exact ground state of Eq.\ (\ref{q1}). The number of such
states is $C_{n-k+1}^{k}$ and they are feasible if
$k<\frac{n+1}{2}$ for OBC. However, the set of states
(\ref{k0-set}) does not present the complete manifold of the
ground states in the sectors of $S_{tot}$ $=S^{z}=S_{\max }-k$ for
$k>2$. Similarly to the two-magnon case we choose the $k$-magnon
set in the form
\begin{equation}
\hat{\Phi} (m_{1})\hat{\Phi} (m_{2})\hat{\Phi} (m_{3})\ldots \hat{\Phi} (m_{k})\left\vert
F\right\rangle ,\quad 1\leq m_{1}<m_{2}<m_{3}<\ldots m_{k}\leq n-1 .
\label{k-set}
\end{equation}
The functions (\ref{k-set}) are exact ground state functions of
the Hamiltonian (\ref{q1}). This can be proved by analogy with the
two-magnon case. We assume again that after projection onto
$S_{tot}=S_{\max }-k$ the set of states (\ref{k-set}) will give a
complete set of linearly independent wave functions in this
sector. As follows from Eq.\ (\ref{k-set}) the number of these
functions is $C_{n-1}^{k}$. Again we have checked and confirmed
this by full  ED for finite delta chains. We note that the
hypothesis about the number of degenerated ground states in the
sector $S_{tot}$ $=S^{z}=S_{\max }-k$ has been suggested in Ref.\
\onlinecite{suzuki} as a guess based on numerical calculations.
The number of functions in Eq.\ (\ref{k-set}) is larger  than the
number of those given in Eq.\ (\ref{k0-set}). Moreover, the
functions of the type described by Eq.\ (\ref{k-set}) are feasible
for any $k$. In particular, for $S_{tot}=\frac{1}{2}$ there is a
single ground state function with zero energy.

In addition to Eq.(\ref{k-set}) we can choose the sets of the
ground state functions in the sectors $S^{z}=S_{\max }-k$ and
$S>S_{\max }-k$. They have the forms
\begin{eqnarray*}
\hat{\Phi} (m_{1})\hat{\Phi} (m_{2})\hat{\Phi} (m_{3})\ldots \hat{\Phi} (m_{k-1})\hat{\Phi} (n)\left\vert
F\right\rangle ,\quad 1 &\leq &m_{1}<m_{2}<m_{3}<\ldots m_{k-1}\leq n-1 \\
\hat{\Phi}(m_{1})\hat{\Phi} (m_{2})\hat{\Phi} (m_{3})\ldots
\hat{\Phi} (m_{k-2})\hat{\Phi} ^{2}(n)\left\vert F\right\rangle
,\quad 1 &\leq &m_{1}<m_{2}<m_{3}<\ldots
m_{k-2}\leq n-1 \\
&&\ldots \\
\hat{\Phi} (m_{1})\hat{\Phi} ^{k-1}(n)\left\vert F\right\rangle ,
&&1 \leq m_1 \leq n-1 \\
\hat{\Phi} ^{k}(n)\left\vert F\right\rangle  &&\quad .
\end{eqnarray*}
This set of functions represents the ground state functions with
$S^{z}=S_{\max}-k$ but $S_{tot}=S_{\max}-k+1$,
$S_{tot}=S_{\max}-k+2$, ..., $S_{tot}=S_{\max}$.

The total number of ground states in the sector $S^{z}=S_{\max}-k$
amounts
\begin{equation}
C_{n-1}^{0}+C_{n-1}^{1}+\ldots +C_{n-1}^{k} .
\end{equation}

Let us now consider the delta chain with PBC. It is evident
that the ground state in the sector $S^{z}=S_{\max }-k$ can be
formed by $k$ non-overlapping localized magnons
\begin{equation}
\hat{\varphi} _{i1}\hat{\varphi} _{i2}
\hat{\varphi} _{i3}\ldots \hat{\varphi} _{ik}\left\vert
F\right\rangle .  \label{k periodic}
\end{equation}
The number of possibilities to place  $k$ magnons on a delta chain
without overlap is
\begin{equation}
g_{n}^{k}=\frac{n}{n-k}C_{n-k}^{k},\quad n=\frac{N}{2}  \label{g} .
\end{equation}
This is the number of degenerated ground states  in the sector
$S^{z}=S_{\max }-k$ built by $k$ non-overlapping localized
magnons. It corresponds to the one-dimensional classical
hard-dimer problem.\cite{fisher,derzhko} The maximum number of
localized magnons for the closest possible packing is $k_{\max
}=\frac{n}{2}$ and $g_{n}^{n/2}=2$. Remarkably, the
non-overlapping localized-magnon states (\ref{k periodic}) do not
exhaust all possible ones for $k>2$. There is another way of the
ground state construction. For example, we can write the exact
ground state for $k=2$ as
\begin{equation}
\hat{\varphi} _{i}(\hat{\varphi} _{i-1}+\hat{\varphi} _{i}+\hat{\varphi} _{i+1})\left\vert
F\right\rangle . \label{p2}
\end{equation}
Carrying out computations similarly to those for the open chain it
is easy to see that the function (\ref{p2}) is an exact
eigenfunction with zero energy for the local Hamiltonians
$\hat{H}_{i}$, $\hat{H}_{i+1}$ and $\hat{H}_{i-1}$ and for the
other ones. Formula (\ref{p2}) can be extended for $k>2$ by adding
corresponding brackets. On the base of the analysis of possible
construction of such type we conjecture that the ground state
degeneracy in the sector $S_{tot}$ $=S^{z}=S_{\max }-k$ amounts
\begin{equation}
A_{n}^{k}=C_{n}^{k}-C_{n}^{k-1}+\delta _{k,n} . \label{An}
\end{equation}
According to Eq.\ (\ref{An}) $A_{n}^{k}=0$ for $n>k>\frac{n}{2}$
and $A_{n}^{n/2}=\frac{2}{2+n}C_{n}^{n/2}$. The third term in
Eq.\ (\ref{An}) corresponds to the special ground state for $S=0$
described by the famous resonating-valence-bond eigenfunction
\cite{hamada} which is not of "multi-magnon" nature. As follows
from Eq.\ (\ref{An}) the number of the ground states for fixed
$S^{z}=S_{\max }-k$ is
\begin{eqnarray}
B_{n}^{k} &=&C_{n}^{k},\quad 0\leq k\leq \frac{n}{2}  \nonumber \\
B_{n}^{k} &=&C_{n}^{n/2}+\delta _{k,n},\quad \frac{n}{2}<k\leq n .
\label{number}
\end{eqnarray}
Eqs.(\ref{An}) and (\ref{number}) have been confirmed by ED
 calculations of finite chains up to $N=24$.

The total number of degenerate ground states is
\begin{equation}
W=2\sum_{k=0}^{n-1}B_{n}^{k}+B_{n}^{n}=2^n+nC_n^{n/2}+1 .
\label{W}
\end{equation}
The value of the entropy per site is $s_0=\ln(W)/N$. That is
the residual entropy per site at zero magnetic field which
becomes  for $N\to\infty$
\begin{equation}
s_0=\frac 12 \ln 2 . \label{s0}
\end{equation}
Obviously, the residual entropy of the considered $N$-site
interacting spin-$1/2$ system corresponds  to the entropy of
$\frac N2$ non-interacting $s=1/2$ spins.
It is interesting to compare the residual entropy of the F-AF
delta chain at the critical point with that for the AF delta chain
at the saturation field. For the AF delta chain it amounts
$s_0^{AF}=0.347\ln 2$ \cite{Zhitomir,derzhko,derz}. i.e.\ $s_0$ is
larger than $s_{AF}$ due to the existence of the additional ground
states which do not belong to the class of non-overlapping
localized magnons.
Concluding this section we point out that the considered model is
one more example of a quantum many-body system with a macroscopic
ground-state degeneracy resulting therefore} in a residual
entropy.

\section{Low-temperature thermodynamics}

The next interesting question is whether the degenerate ground
states are separated by a finite gap from all other eigenstates.
This question is important for thermodynamic properties of the
model. If a finite gap exists in all spin sectors then the
low-temperature thermodynamics is determined by the contribution
of the degenerate ground states. Such a situation takes place for
the delta chain with antiferromagnetic interactions. As it will be
demonstrated below it is not the case for the considered model.

As follows from Eq.(\ref{one-magnon}) the gap $\Delta E$ in the
one-magnon sector is $\Delta E=1$ (in $\left\vert J_{1}\right\vert
$ units). However, the minimal energy of two-magnon excitations
dramatically decreases. Numerical calculations show that it equals
$\Delta E\approx $ $0.022$. The exact wave function of this state
has the form
\begin{eqnarray}
\Psi
&=&0.484\sum_{n}(-1)^{n}s_{2n}^{-}(s_{2n-1}^{-}+s_{2n+1}^{-})\left\vert
F\right\rangle \nonumber \\
&&-0.321\sum_{n}\sum_{m=0}(-1)^{n}\exp (-\lambda
m)s_{2n}^{-}(s_{2n-2m-3}^{-}+s_{2n+2m+3}^{-})\left\vert
F\right\rangle
\nonumber \\
&&+0.545\sum_{n}\sum_{m=1}(-1)^{n}\exp \{-\lambda
(m-1)\}s_{2n+1}^{-}s_{2n+4m-1}^{-}\left\vert F\right\rangle \nonumber \\
&&-0.157\sum_{n}\sum_{m=0}(-1)^{n}\exp (-\lambda
m)s_{2n}^{-}s_{2n+4m}^{-}\left\vert F\right\rangle ,
\label{two-magnon gap}
\end{eqnarray}%
where $\lambda \simeq 3.494$. The energy of this state is $\Delta
E=0.02177676$. It could be expected that the low-lying excited
two-magnon states are formed by scattering states of magnons from
the dispersionless one-magnon branch. However, the wave function
(\ref{two-magnon gap}) has a more complicated specific form of a
bound state.

The gaps for the $k$-magnon states with $k>2$ decrease rapidly
with increasing $k$ as it can be seen from the Table 1, where the
gaps in the sector $S=S_{\max }-k$ for chains with $N=16,20,24$
are presented. Obviously, the gaps become extremely small.

\begin{table}[tbp]
\caption{Excitation gaps in the $k$-magnon sectors (i.e. $S_z=N/2
- k$) calculated for $N=16,20,24$.}
\begin{ruledtabular}
\begin{tabular}{cccc}
& $N=16$ & $N=20$ & $N=24$ \\
\hline
$k=1$ & $1.0$ & $1.0$ & $1.0$ \\
$k=2$ & $0.021776237324972$ & $0.021776745369208$ & $0.021776760796279$ \\
$k=3$ & $0.000471848035563$ & $0.000484876324415$ & $0.000487488767250$ \\
$k=4$ & $0.000009935109570$ & $0.000013213815119$ & $0.000014315249351$ \\
$k=5$ & $0.000003034124289$ & $0.000000197371592$ & $0.000000295115215$ \\
$k=6$ & $0.000002583642491$ & $0.000000064146143$ & $0.000000004288885$ \\
\end{tabular}
\end{ruledtabular}
\end{table}

These data clearly testify that the contribution of the excited
states to the partition function cannot be neglected even for very
low temperatures. Nevertheless, to clarify this point it is proper
to calculate the contribution to the partition function from only
the degenerate ground states. Using Eq.\ (\ref{number}) we obtain
the partition function $Z$ of the model in the magnetic field in a
form (we use PBC for the calculation since $Z$ for the chains with
PBC and OBC coincide in the thermodynamic limit)
\begin{equation}
Z=2\sum_{k=0}^{n/2}C_{n}^{k}\cosh \left[ \frac{(n-k)h}{T}\right]
+2C_{n}^{n/2}\sum_{k=0}^{n/2}\cosh \left[ \frac{(\frac{n}{2}-k)h}{T}\right]
-2C_{n}^{n/2}\cosh \left( \frac{nh}{2T}\right) -C_{n}^{n/2} . \label{Z}
\end{equation}
The magnetization is given by
\begin{equation}
M=\left\langle S^{z}\right\rangle =T\frac{d\ln Z}{dh} .  \label{M}
\end{equation}
It follows from Eqs.~(\ref{Z}) and (\ref{M}) that $M$ is a
function of the universal variable $x=h/T$. The dependence $M(x)$
is shown in Fig.\ 2 for different $N$. As it is seen from Fig.\ 2
for small $x$ the magnetization grows with the increase of $N$.
Analyzing the magnetization curve $M(x)$ for small $x$ one needs
to distinguish the limits $x\ll 1/N$ and $x\gg 1/N$. Using Eqs.\
(\ref{Z}) and (\ref{M}) we obtain the magnetization for $x\ll 1/N$
in the form

\begin{figure}[tbp]
\includegraphics[width=5in,angle=0]{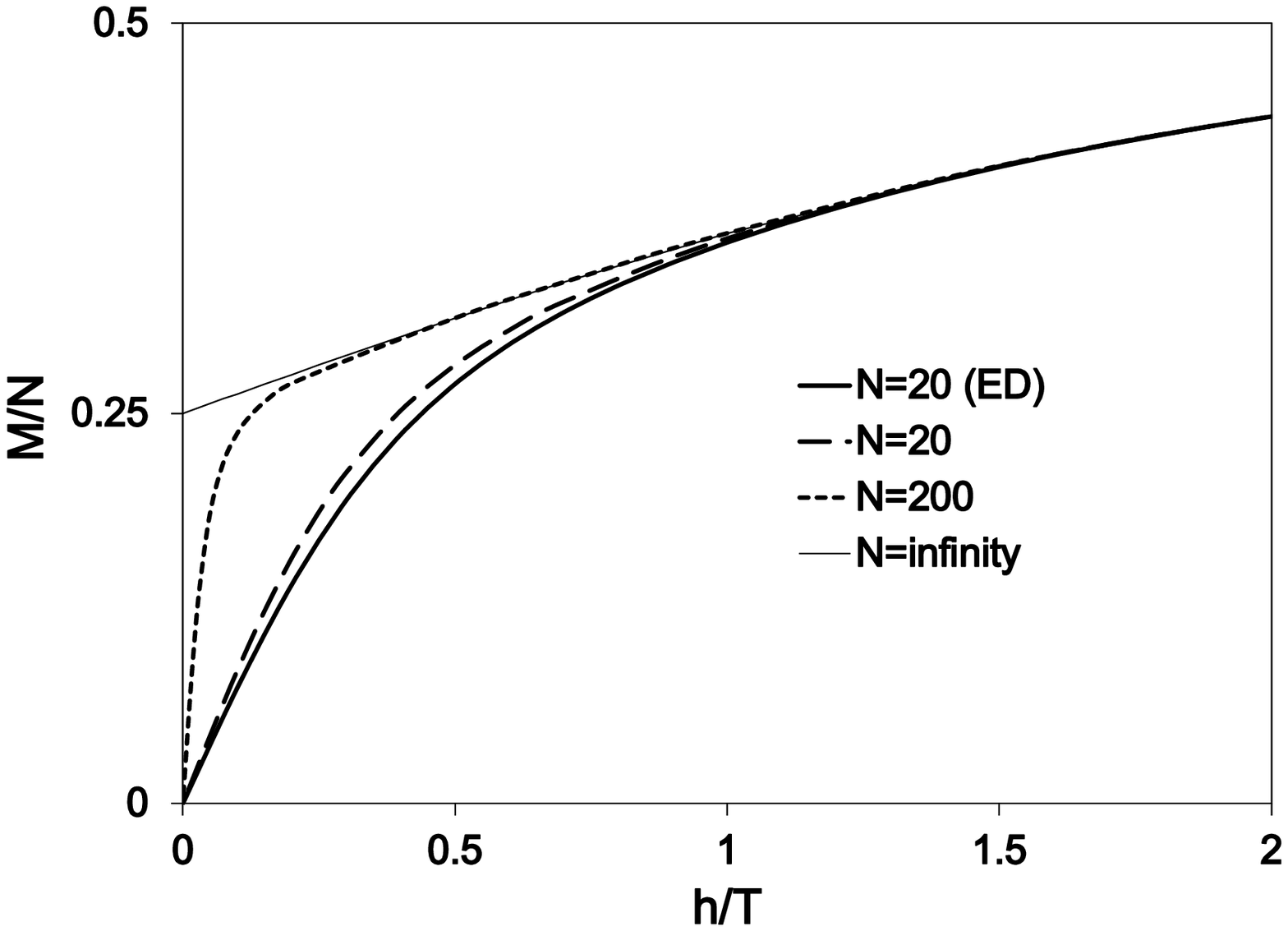}
\caption{Magnetization curves calculated using Eqs.~(\ref{Z}) and
(\ref{M}) for $N=20$ (long-dashed line), $N=200$ (short-dashed
line) and using Eq.(\ref{M2}) for $N\to\infty$ (thin solid line).
Thick solid line corresponds to ED for $N=20$
and $T=10^{-6}$.} \label{M_hgs}
\end{figure}

\begin{equation}
M=c_{N}\frac{N^2h}{T},\quad c_{N}=\frac{2^{n-2}n(n+1)+C_{n}^{n/2}(\frac{3}{4}%
n^{2} +\frac{1}{2}C_{n}^{3})}{n^{2}2^{n+2}+4n^{3}C_{n}^{n/2}} .
\label{M1}
\end{equation}
For $N\gg 1$, $c_{N}\sim 1/48$ and the magnetization per site
becomes
\begin{equation}
\frac{M}{N}\simeq \frac{Nh}{48T}(1+2\sqrt{\frac{\pi }{N}}),\quad h\ll T/N .
\label{M11}
\end{equation}
In the opposite limit $x\gg 1/N$, the magnetization is
\begin{equation}
\frac{M}{N}\simeq \frac{1}{2(1+e^{-h/T})},\quad h\gg T/N .
\label{M2}
\end{equation}
However, it is clear that both equations (\ref{M11}) and
(\ref{M2}) do not give an adequate description of the
magnetization at $x\to 0$. For $x\ll 1/N $, $M$ is proportional to
$N^{2}$ instead of to $N$. On the other hand, according to Eq.\
(\ref{M2}), the magnetization in the thermodynamic limit is finite
at $h=0$. This is an artefact because the long range order (the
magnetization) in one-dimensional systems can not exist at $T>0$.
Therefore, the contribution of only the degenerate ground states
is not sufficient to describe the correct dependence of $M(x)$ for
small $x$ and it is necessary to take into account the
contributions of other low-lying eigenstates. Unfortunately,
analytical calculation of the corresponding contributions is
impossible. Therefore, we carried out the full ED for $N=16$ and
$N=20$.

The magnetization curves obtained by ED calculations are shown in
Fig.~3. It is seen that curves for $N=16$ and $N=20$ are close
(especially at $h/T>1$) that testifies small finite-size effects.
One of the most interesting points related to the magnetization
curve is its behavior at low magnetic fields. At first, we note
that $M$ obtained by ED calculations is not a function of only
$x=h/T$ in contrast with the predictions given by
Eqs.~(\ref{M11}), (\ref{M2}). That can be seen in the inset in
Fig.~3, where the magnetization for $N=20$ is presented as a
function of $x$ for two temperatures, $T=10^{-4}$ and $T=10^{-5}$,
i.e. in fact, $M=M(x,T)$.

\begin{figure}[tbp]
\includegraphics[width=5in,angle=0]{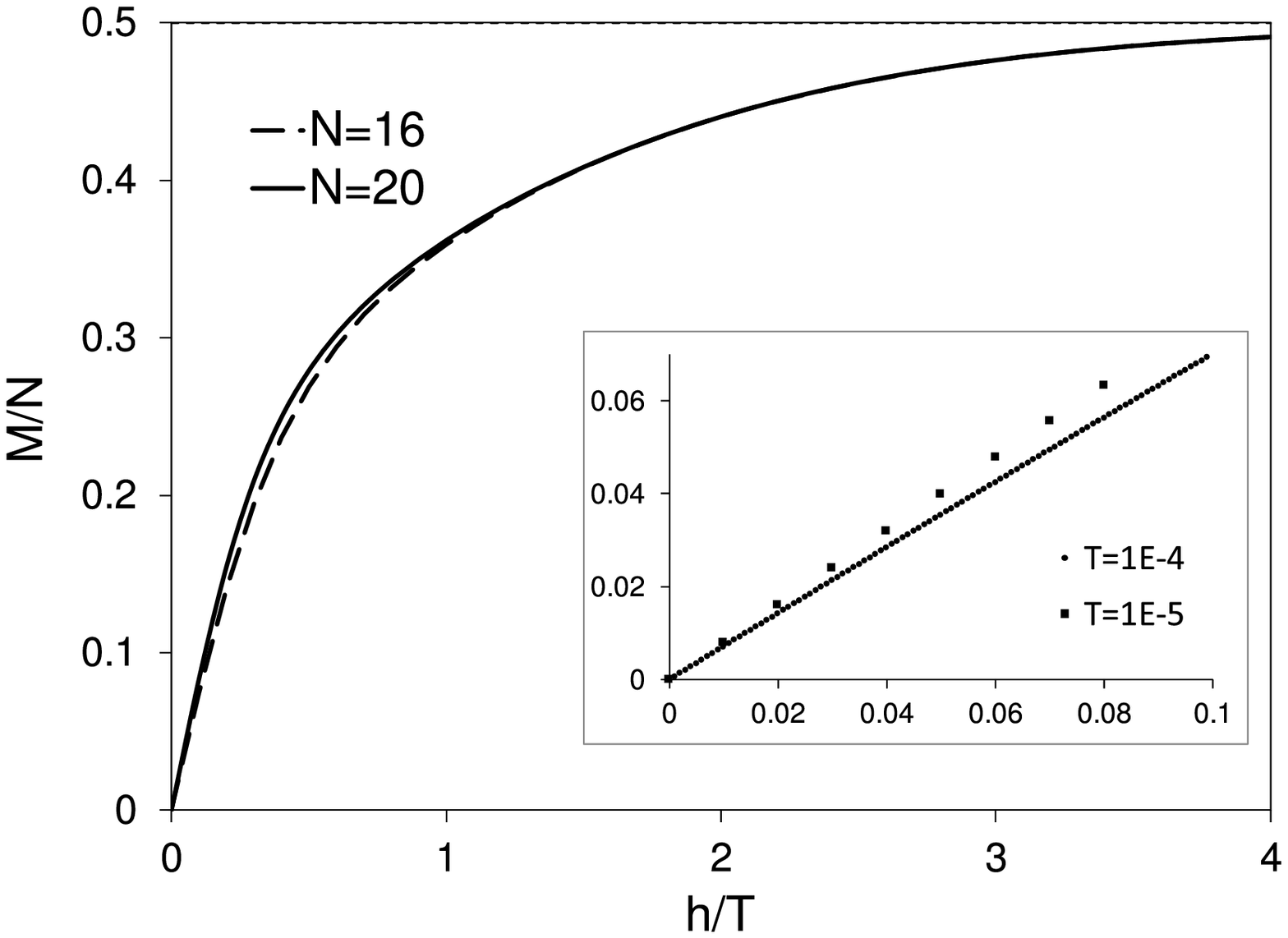}
\caption{Magnetization curves calculated by ED
for $N=16$ and $N=20$ at fixed temperature $T=10^{-6}$. The inset
shows low-field limit of the magnetization curve calculated for
$N=20$ and two temperatures $T=10^{-4}$ and $T=10^{-5}$.}
\label{M_h}
\end{figure}

In order to study the low-field limit of the magnetization curve
we have calculated the uniform susceptibility per site
\begin{equation}
\chi =\frac{1}{3NT}\sum_{ij}\left\langle \mathbf{S}_{i}\cdot \mathbf{S}%
_{j}\right\rangle  .\label{cor}
\end{equation}
The calculated dependencies of $\chi (T)$ for $N=16$ and $N=20$
are shown in Fig.\ 4. For convenience they are plotted as
$\ln(\chi T)$ vs.\ $\ln T$. Both curves are almost
indistinguishable for $T>10^{-3}$, indicating a weak finite-size
dependence. A linear fit in this temperature range for the log-log
plot of $\chi (T)$ yields a power-law dependence
\begin{equation}
\chi =\frac{c_{\chi }}{T^{\alpha }}  \label{sus}
\end{equation}
with
\begin{eqnarray}
c_{\chi } &\simeq &0.317  \nonumber \\
\alpha  &\simeq &1.09  \label{alpha}
\end{eqnarray}

\begin{figure}[tbp]
\includegraphics[width=5in,angle=0]{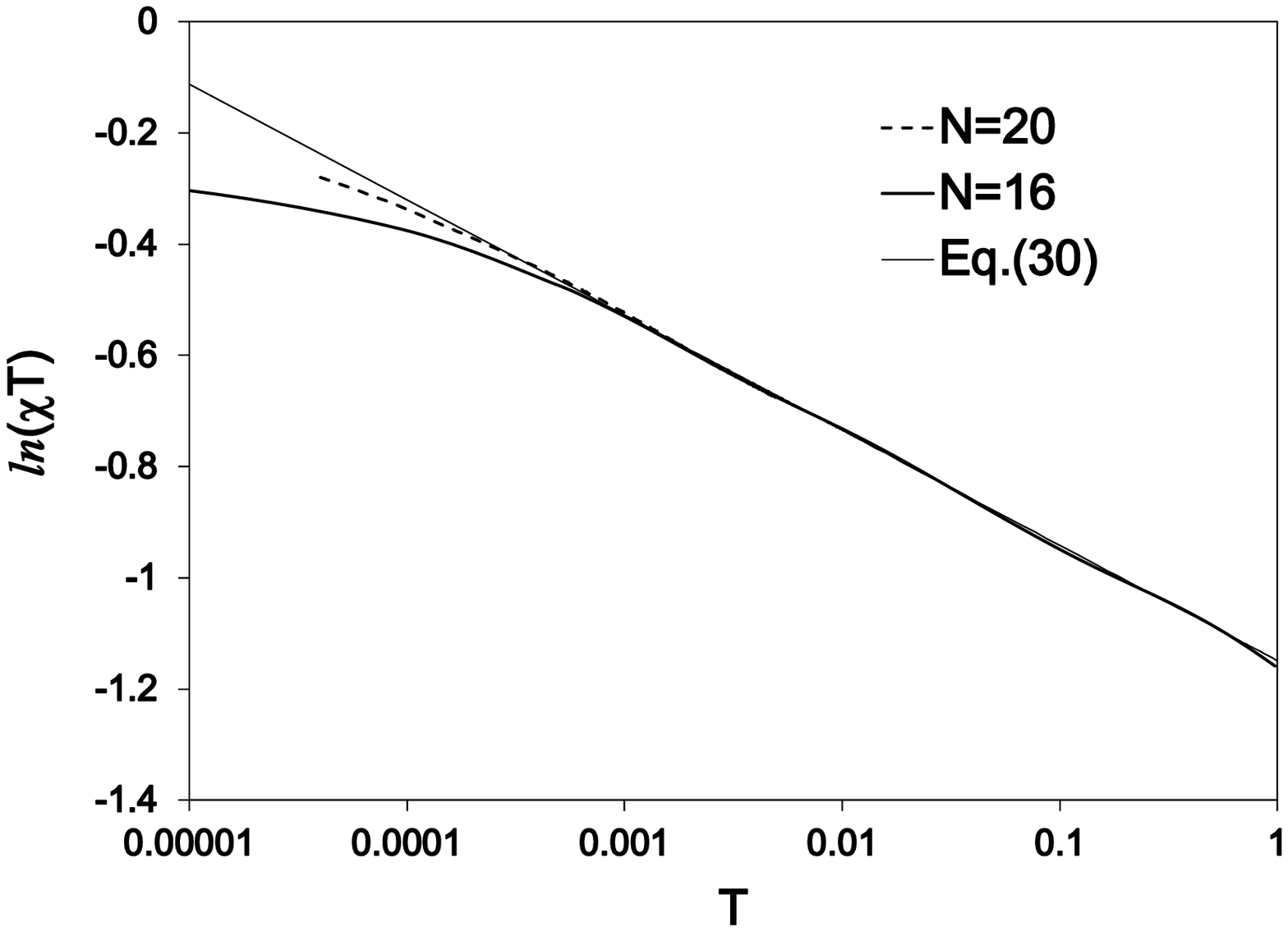}
\caption{Log-log plot for the dependence of the susceptibility per
site on temperature calculated for $N=16$ and $N=20$. The thin
solid line corresponds to Eq.\ (\ref{sus}).} \label{lnchi_lnT}
\end{figure}

As shown in Fig.\ 4, Eq.\ (\ref{sus}) perfectly coincides with the
numerical data for $N=16$ and $N=20$ from $T\sim 10^{-3}$ up to
$T=1$, only slight deviations near $T=0.1$ and $T=1$ are observed.
However, for $T<10^{-3}$ the curves $\chi(T)$ for $N=16$ and
$N=20$ start to split and both deviate from Eq.\ (\ref{sus}).

At $T\to 0$ the susceptibility is determined by the contribution
of the degenerate ground states and it is
\begin{equation}
\chi =c_N\frac{N}{T}.  \label{susT0}
\end{equation}
with $c_{N}$ given by Eq.(\ref{M1}). For $N\gg 1$ it reduces to
$\chi =N/48T$.

We assume that both expressions for the susceptibility (\ref{sus})
and (\ref{susT0}) are described by a single universal finite-size
scaling function. This guess leads to the following form for the
finite-size  susceptibility:
\begin{equation}
\chi_{N}(T)=T^{-\alpha }f(c_{N}NT^{\alpha -1})  \label{chiscal}
\end{equation}
Really, the behavior of the scaling function $f(z)=z$ for $z\ll 1$
provides the correct limit to Eq.\ (\ref{susT0}). In the
thermodynamic limit when $z=c_{N}NT^{\alpha -1}\to\infty$ the
scaling function $f(z)$ tends to a finite value $c_{\chi }$ in
full accord with Eq.\ (\ref{sus}). The crossover between the two
types of the susceptibility behavior occurs at $z\sim 1 $, which
defines the effective temperature of the crossover $T_{0}\sim
N^{-1/(\alpha -1)}$. At $T<T_{0}$ the susceptibility is determined
mainly by the contribution of the degenerate ground states, but
this regime vanishes in the thermodynamic limit where $T_{0}=0$.
Substituting the value $\alpha \simeq 1.09$ we obtain a very large
exponent $\simeq 11$ for $T_{0}\sim 1/N^{11} $. This exponent
defines the energy scale of the excited states which contribute to
the susceptibility.

\begin{figure}[tbp]
\includegraphics[width=5in,angle=0]{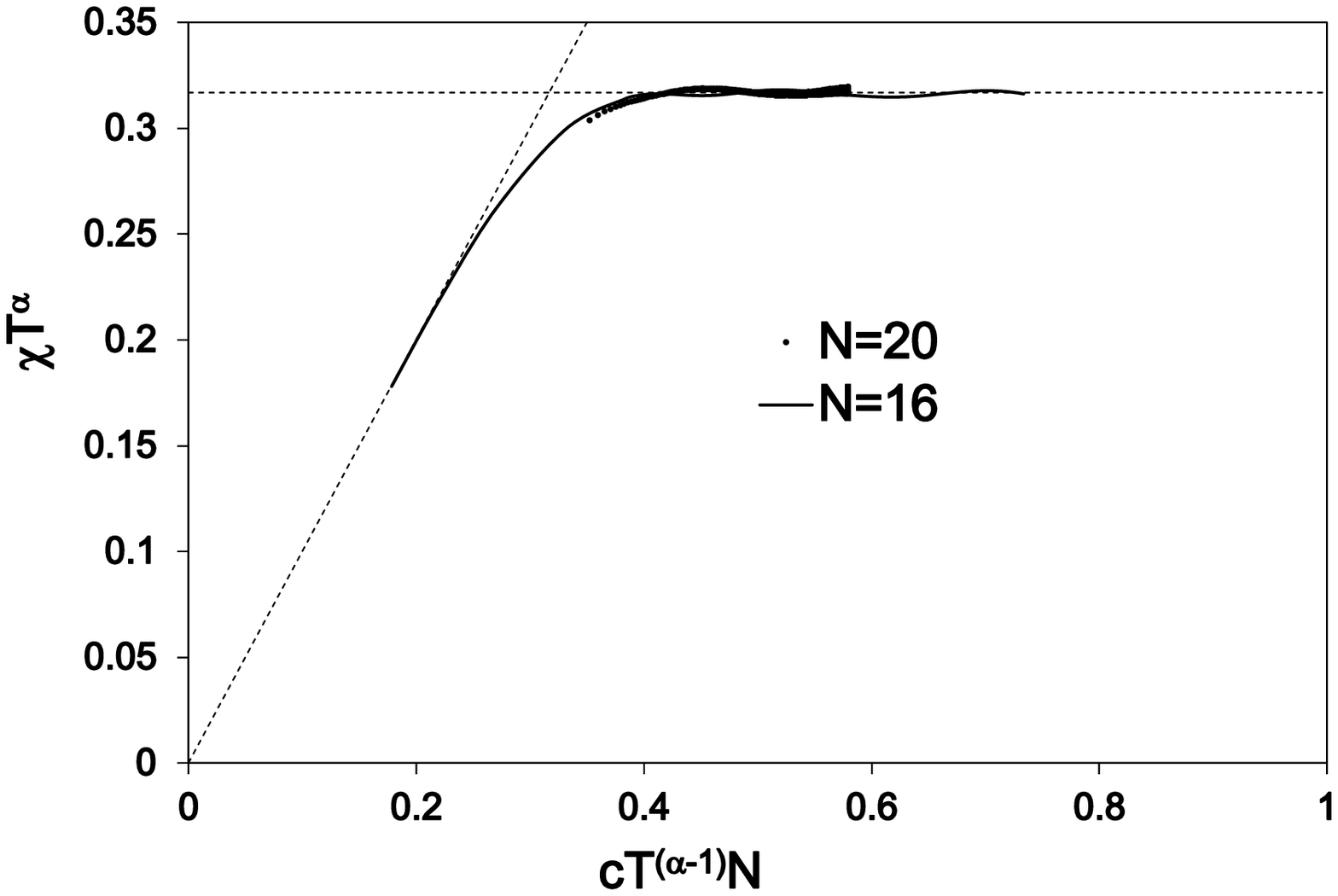}
\caption{Universal scaling function for the dependence of the
finite-size susceptibility on temperature defined in
Eq.(\ref{chiscal}) calculated by ED for $N=16$ and $N=20$. Thin
dashed lines correspond to Eqs.\ (\ref{sus}) and (\ref{susT0}).}
\label{Fig_scale}
\end{figure}

The scaling hypothesis written in Eq.\ (\ref{chiscal}) is
confirmed numerically. In Fig.\ \ref{Fig_scale} the ED data for
$N=16$ and $N=20$ are plotted in the axes $\chi_N T^{\alpha }$
vs.\ $c_{N}NT^{\alpha -1}$. As shown in Fig.\ \ref{Fig_scale} the
data for $N=16$ and $N=20$ lie very close and define the scaling
function $f(z)$.

The obtained temperature dependence $\chi (T)$ (\ref{sus}) allows
us to determine the low-field behavior of the magnetization curve
\begin{equation}
\frac{M}{N}=c_{\chi }\frac{h}{T^{\alpha }} .  \label{Mlowh}
\end{equation}
This implies that the low field magnetization is a function of a
single scaling variable $y=h/T^{\alpha }$. This statement is
confirmed by numerical calculations, presented in Fig.\ 6. As
shown in Fig.\ 6 the magnetization calculated for different (and
small) values of the field $h$ and the temperature $T$ lies on one
line when it is plotted against the scaling variable
$y=h/T^{\alpha }$ with $\alpha =1.09$.

\begin{figure}[tbp]
\includegraphics[width=5in,angle=0]{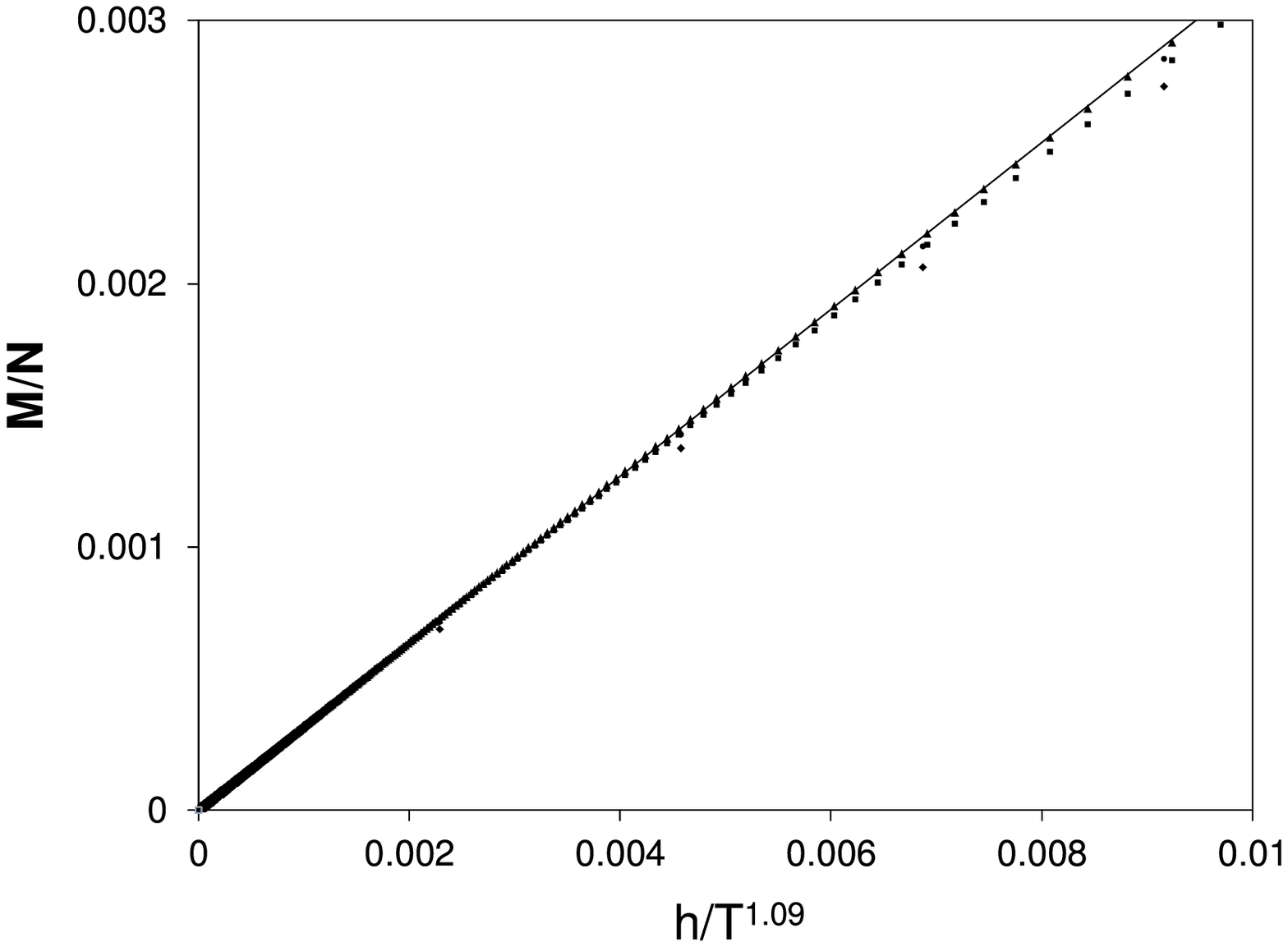}
\caption{Dependence of the magnetization per site on the scaling
parameter $y=h/T^{1.09}$ calculated by ED ($N=20$) for different
values of the magnetic field $h$ and temperature $T$. Thin solid
line corresponds to Eq.\ (\ref{Mlowh}).} \label{M_lowh109}
\end{figure}

The temperature dependence of the spin correlation functions
$\langle {\bf S}_i\cdot{\bf S}_i\rangle $ for $N=16$ is presented
in Fig.\ 7. For low temperature up to $T\leq 10^{-3}$ the spin
correlation functions are almost constants and the sum in Eq.\
(\ref{cor}) at $T=10^{-9}$ is equal to $c_{16}$ with $c_{16}$
given by Eq.\ (\ref{M1}). For $T>10^{-3}$ the correlations decay
with the increase of $T$ and with the distance between the spins.

\begin{figure}[tbp]
\includegraphics[width=5in,angle=0]{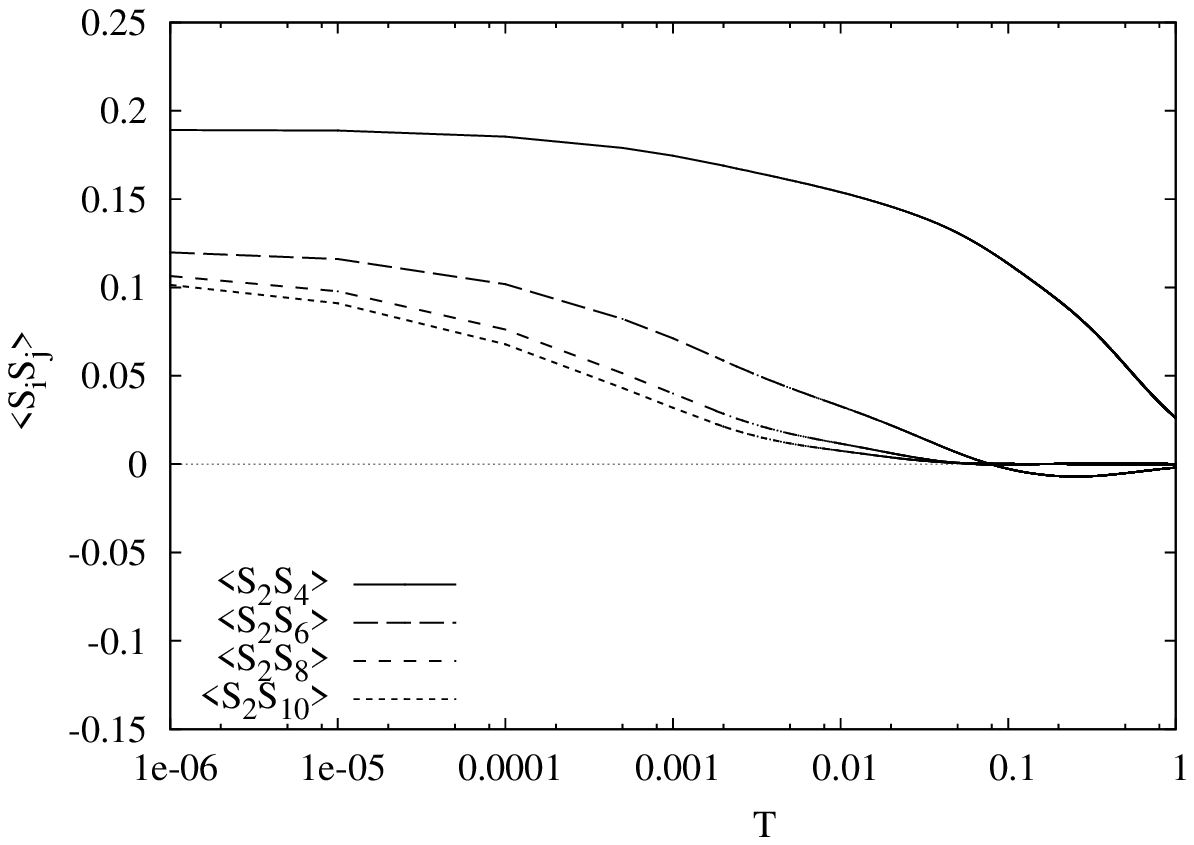}
\caption{Temperature dependence of various spin correlators
$\langle {\bf S}_i\cdot{\bf S}_i\rangle $  (ED data for $N=16$.)
The numbering in the legend corresponds to Fig.~\ref{fig1}
(periodic boundary conditions imposed).} \label{sisj}
\end{figure}

Let us consider now the entropy and the specific heat. We note
that the partition function (\ref{Z}) at $h=0$ does not depend on
the temperature, and the Helmholtz free energy is
\begin{equation}
\frac{F}{N}=-T\ln Z=-TS_{0}
\end{equation}
The fact that $Z$ in Eq.\ (\ref{Z}) does not depend on $T$ at $h=0$ means that
the partition function (\ref{Z}) is not relevant at $T>0$.
Nevertheless, Eq.\ (\ref{Z}) gives the exact value for the residual
entropy given by Eqs.\ (\ref{W}) and (\ref{s0}).

The numerical data for the $T$-dependence of the entropy at $h=0$
obtained by ED are shown in Fig.\ 8. As it is there, the data for
$N=16$ and $N=20$ perfectly coincide for $T>10^{-3}$ and split for
$T<10^{-3}$. At $T\to 0$ the entropy for $N=16$ and $N=20$ tends
to different values of the residual value given by Eq.\ (\ref{W}).
From these facts we conclude that the finite-size effects in our
calculations become substantial for $T<10^{-3}$, but the obtained
data for $T>10^{-3}$ perfectly describes the behavior of the
entropy at $N\to \infty $. Therefore, we used the data for
$T>10^{-3}$ only, and found that the behavior of the entropy in
the thermodynamic limit is to first approximation reasonably well
described by a power-law dependence (see Fig.8):
\begin{equation}
\frac{S(T)}{N}=\frac{1}{2}\ln 2+c_{s}T^{\lambda } \label{STfit}
\end{equation}
with $c_{s}\simeq 0.245$ and $\lambda \simeq 0.12$.

\begin{figure}[tbp]
\includegraphics[width=5in,angle=0]{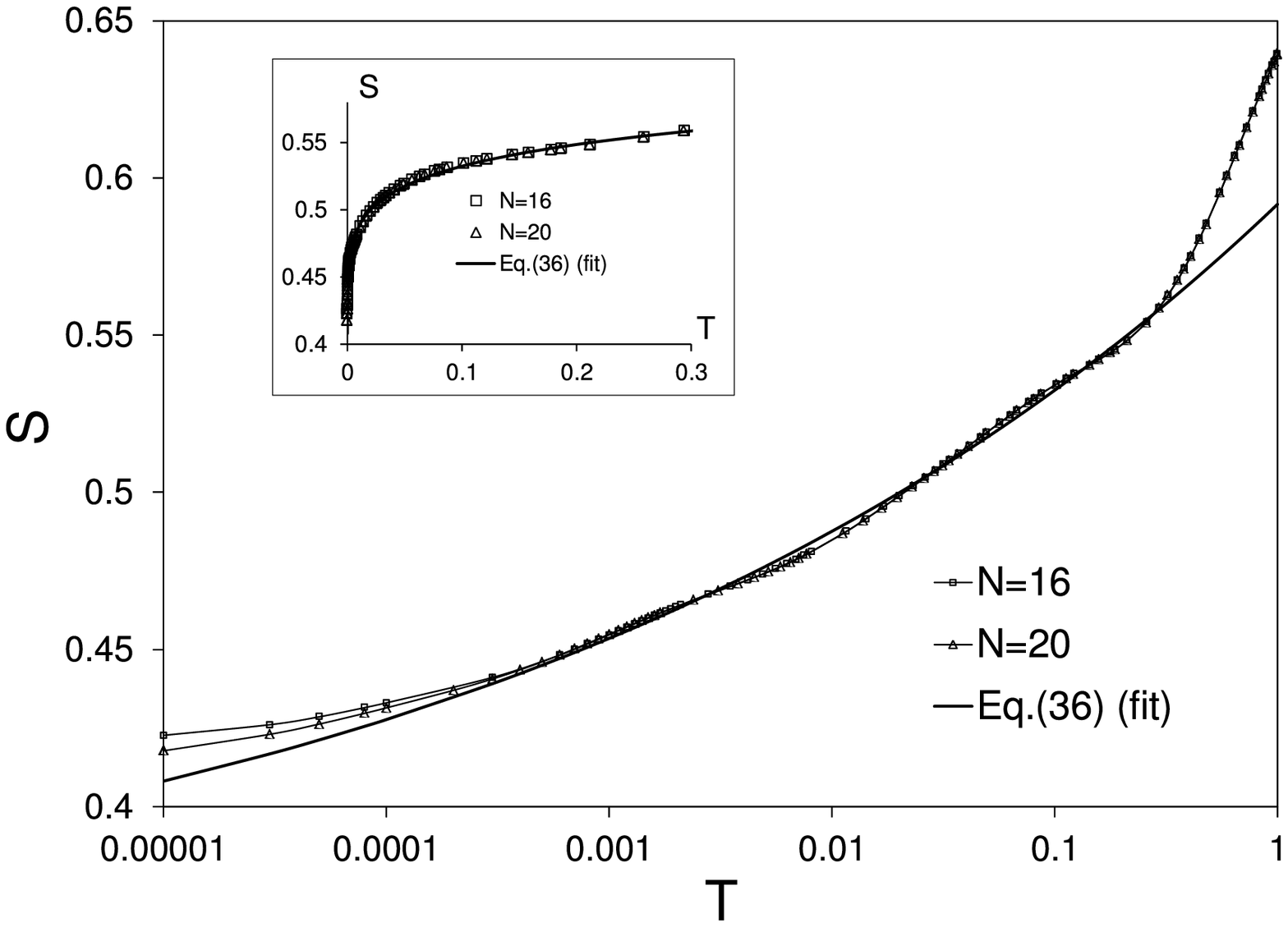}
\caption{Dependence of the entropy per site on temperature
calculated for $N=16$ and $N=20$ and presented in a logarithmic
scale. The thick solid line describes the approximate smooth
expression given by Eq.\ (\ref{STfit}). The inset shows the
low-temperature limit of $S(T)$.} \label{entropy_fig}
\end{figure}

The dependence of the specific heat on the temperature is
presented in Fig.\ 9. It has a peculiar form and is characterized by
a broad maximum at $T\simeq 0.7$ and two weak maxima at $T\leq
0.1$.

It is important to note that the data for $N=16$ and $N=20$ are
slightly different at $T<10^{-3}$ but they are indistinguishable
for $T>10^{-3}$, testifying to these data are already close to
those for the thermodynamic limit. Therefore, we conclude that the
prominent feature of this dependence remains relevant at $N\to
\infty$.

\begin{figure}[tbp]
\includegraphics[width=5in,angle=0]{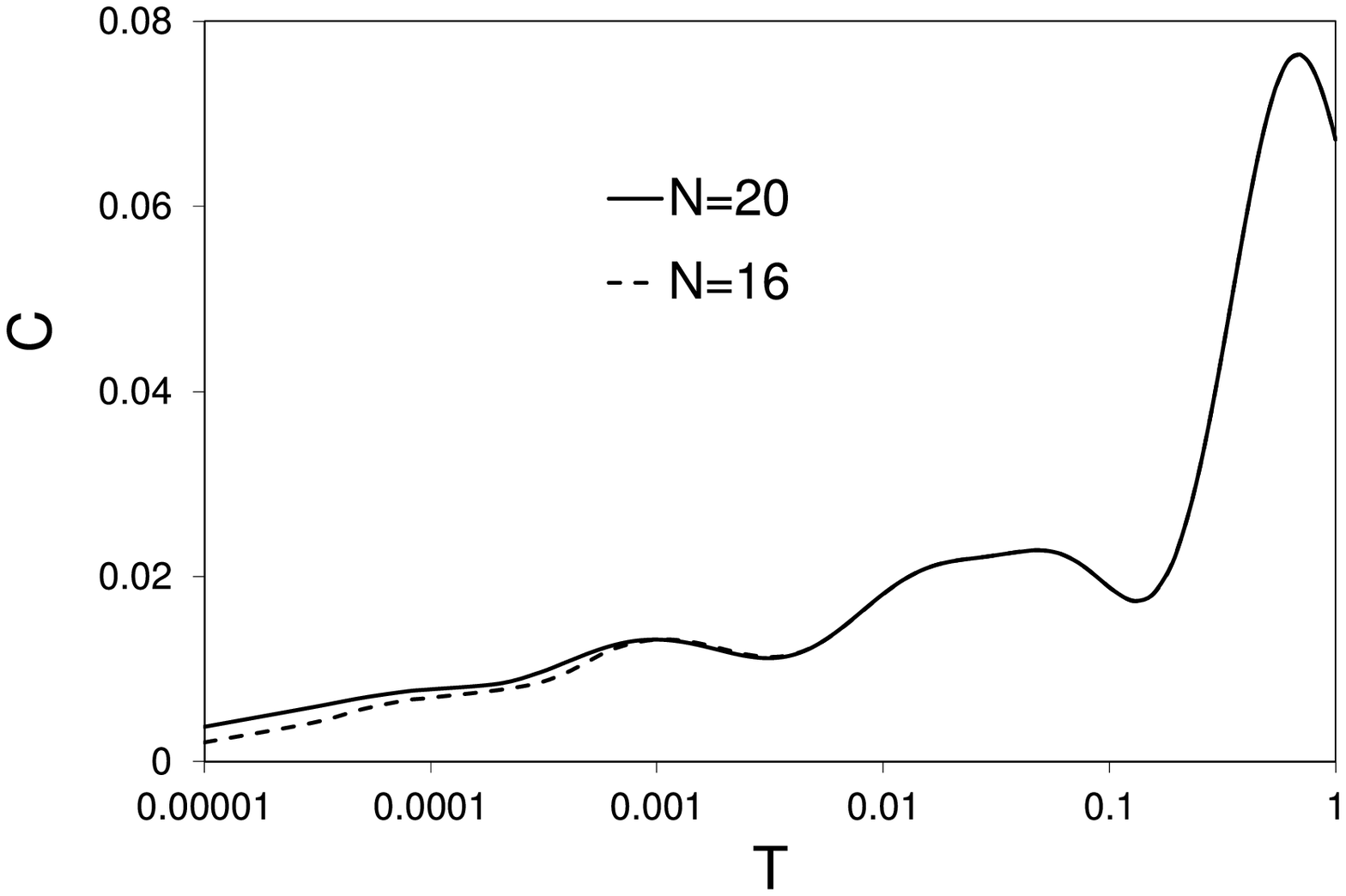}
\caption{Dependence of the specific heat on temperature calculated for
$N=16$ (dashed line) and $N=20$ (solid line).} \label{C_T}
\end{figure}

\section{Magnetocaloric effect}

As it is well-known \cite{Zhit} that spin systems with a
macroscopic degenerate ground state show an appreciable
magnetocaloric effect, i.e.\ for the cooling of the system under
an adiabatic demagnetization. The standard materials for magnetic
cooling are paramagnetic salts. The geometrically frustrated
quantum spin systems can be considered as alternative materials
for low-temperature magnetic cooling. The macroscopic degeneracy
of the ground state at the saturation magnetic field in some of
them, including the AF delta chain, leads to an enhanced
magnetocaloric effect in the vicinity of this field
\cite{Honecker,Zhit1,derzhko2006,Schmidt,Garlatti}. However, the
saturation field is relatively high in real materials and
practical applications of such systems for magnetic cooling are
rather questionable.

In contrast, the F-AF delta chain with $\alpha =\frac{1}{2}$ has a
finite zero-temperature entropy at zero magnetic field. Therefore,
it is interesting to consider the magnetocaloric properties of
this model. The efficiency of the magnetic cooling is
characterized by the cooling rate $(\frac{\partial T}{\partial
h})_{s}$ and so it is determined by the dependence $T(h)$ at a
fixed value of the entropy. This dependence at small $h$ and $T$
can be found using the results obtained in the previous Sections.
According to the standard thermodynamic relations the entropy
$S(T,h)$ is connected with the magnetization curve by
\begin{equation}
S(T,h)-S(T,0)=\frac{\partial }{\partial T}\int_{0}^{h}M(T,h')dh'
\label{entropy2}
\end{equation}
As was stated in the previous Section, there are two regions with
different behavior of the magnetization curve. For very low
magnetic field $h<T^{\alpha}$ the magnetization is proportional to
$h$ according to Eq.\ (\ref{Mlowh}). For higher magnetic field
$h>T^{\alpha}$ (but both $h\ll 1$ and $T\ll 1$) the magnetization
curve is described by Eq.\ (\ref{M2}). Therefore, we will consider
these two cases separately.

At first we study the low-field case $h<T^{\alpha}$. Substituting
the expression (\ref{Mlowh}) to Eq.\ (\ref{entropy2}) we obtain the
entropy per site $s(T,h)=S(T,h)/N$:
\begin{equation}
s(T,h)=s(T,0)-\frac{\alpha c_{\chi }h^{2}}{2T^{\alpha +1}}
\label{entropy1}
\end{equation}
where the function $s(T,0)=S(T,0)/N$ is given by Eq.\ (\ref{STfit}).
>From Eq.\ (\ref{entropy1}) we obtain the function $h(T)$ at
constant entropy $s(T,h)=s^{\ast }$ as
\begin{equation}
h(T)=\sqrt{\frac{2(s_{0}+c_{s}T^{\lambda }-s^{\ast })}{\alpha
c_{\chi }}}T^{(\alpha +1)/2} \label{hT1}
\end{equation}
where $s_0=\ln 2/2$ as given by Eq.\ (\ref{s0}).
From Eq.\ (\ref{hT1}) we see that the cases $s^{\ast}<s_{0}$ and
$s^{\ast}>s_{0}$ are different. For the case $s^{\ast}\geq s_{0}$
the temperature tends to the finite value $T_{0}$ at $h\to 0$:
\begin{equation}
T_{0}=\left( \frac{s^{\ast }-s_{0}}{c_{s}} \right)^{1/\lambda } .
\label{T0}
\end{equation}

In other words $T_{0}$ is the lowest temperature which can be
reached in the adiabatic demagnetization process if the entropy
exceeds $s_{0}$. For low magnetic fields Eq.\ (\ref{hT1}) allows
to express the dependence $T(h)$ as:
\begin{equation}
T(h)=T_{0}+\frac{\alpha c_{\chi} h^{2}}{2\lambda
c_{s}T_{0}^{\alpha +\lambda }} .
\end{equation}
In the limit $T\gg T_{0}$, the curve $T(h)$ transforms into
\begin{equation}
T(h)=\left(\frac{\alpha c_{\chi }}{2c_{s}}\right)^{1/(1+\alpha
+\lambda )}h^{2/(1+\alpha +\lambda )} .
\end{equation}
Substituting the values for $\alpha$, $c_{\chi }$, $\lambda$ and
$c_{s}$ into the latter equation, we get
\begin{equation}
T(h)\simeq 0.85 h^{0.905} \label{Th1}
\end{equation}
which gives the cooling rate
\begin{equation}
\left(\frac{\partial T}{\partial h}\right)_{s^{\ast}} \simeq 0.77
h^{-0.095} . \label{dTdh1}
\end{equation}
As follows from Eq.\ (\ref{T0}) for the special case
$s^{\ast}=s_{0}$ the critical temperature $T_0=0$ and
Eqs.(\ref{Th1}) and (\ref{dTdh1}) are valid in the low temperature
limit.

In the case $s^{\ast }<s_{0}$ we can omit the term
$c_{s}T^{\lambda }$ in Eq.(\ref{hT1}), which means that $T\to 0$
at $h\to 0$. The cooling rate for $T\ll (s_{0}-s^{\ast
})^{1/\lambda }$ is given by the following expression:
\begin{equation}
\left(\frac{\partial T}{\partial h}\right)_{s^{\ast}}=
\frac{0.413}{(s_{0}-s^{\ast })^{0.48}}h^{-0.043} .
\end{equation}
For the case of small $h$ and $T$ but $h/T\gg 1$ we can calculate
the integral in Eq.\ (\ref{entropy2}) using the expression for the
magnetization given by Eq.\ (\ref{M2}). Then the entropy $s^{\ast }$
is
\begin{equation}
s^{\ast }=\frac{1}{2}\ln (1+e^{-h/T})+\frac{h}{2T(e^{h/T}+1)} .
\label{entropy3}
\end{equation}
This entropy coincides with the entropy per site of the ideal
paramagnet of $\frac{N}{2}$ spins $\frac{1}{2}$. The
transcendental Eq.\ (\ref{entropy3}) does not allow to derive an
explicit expression for $T(h)$. However, since the magnetic field
and the temperature enter Eq.\ (\ref{entropy3}) only in the
combination $h/T$, the dependence $T(h)$ is a linear function. In
the limit $h/T\gg 1$ ($s^{\ast }\ll 1$) one has $T(h)\sim -h/\ln
(2s^{\ast })$.

We have calculated the function $T(h)$ by ED for $N=16$ for
several fixed values of the entropy, see  Fig.\ 10. It is seen
there that the cooling rate increases when $s^{\ast}$ approaches
$s_{0}$ from below. For $s^{\ast}>s_{0}$ a nonzero $T_{0}$
appears, but for $T>T_{0}$ the cooling rate is rather high. For
small $h$ and $T$ the behavior of the curves $T(h)$ agrees with
that given by Eqs.\ (\ref{entropy2})-(\ref{entropy3}).

Having in mind real materials for applications one should be aware
that the expected magnetocaloric effect is expected to be somewhat
reduced due to deviations from the critical point considered here
and always present residual interactions beyond those considered
in Eq.\ (1). A quantitative and systematic study of these cases is
postponed to subsequent studies.

\begin{figure}[tbp]
\includegraphics[width=5in,angle=0]{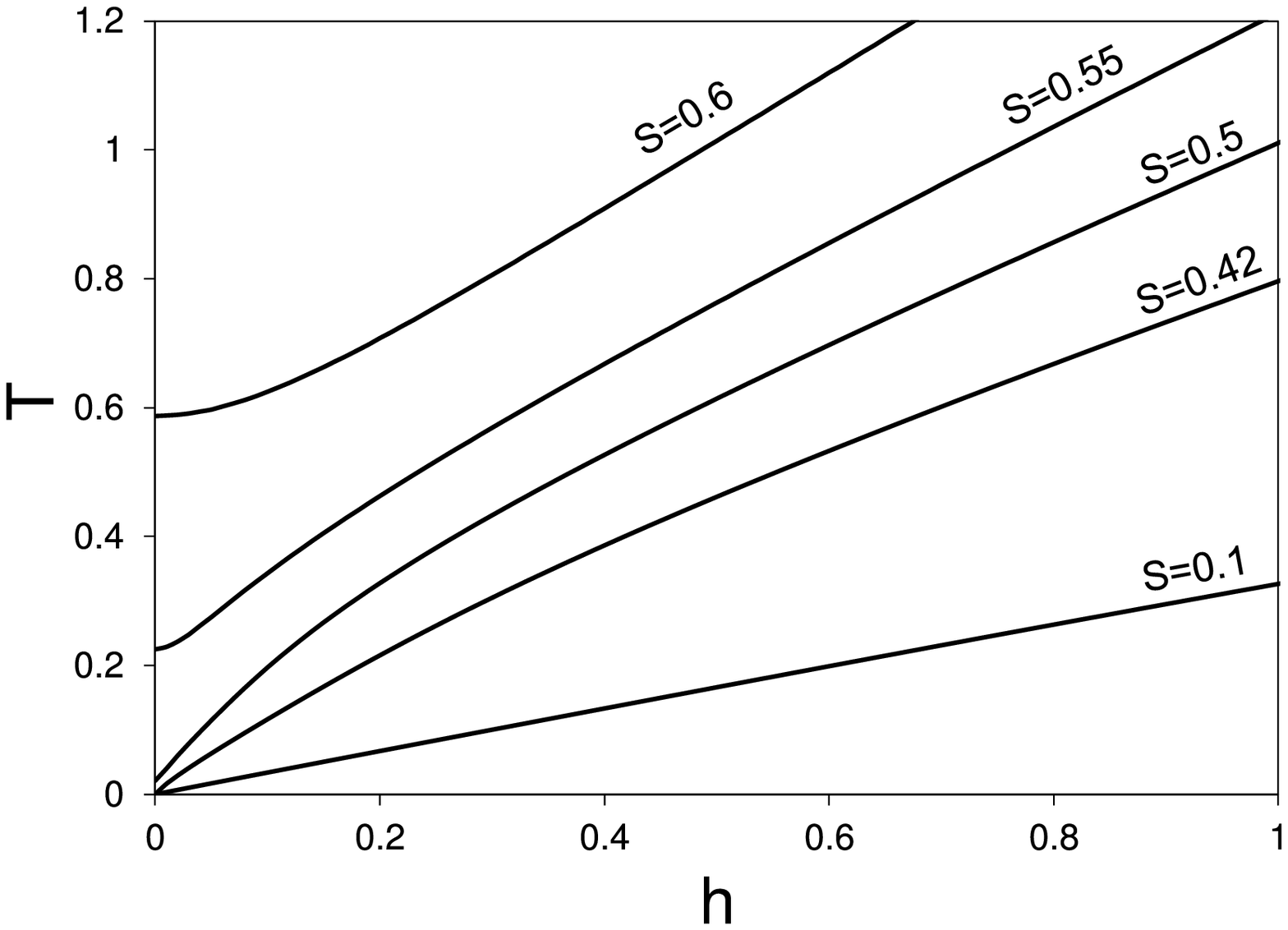}
\caption{Constant entropy curves as a function of the applied
magnetic field and temperature for $N=16$.} \label{calorfig}
\end{figure}

\section{Conclusion}

We have studied the ground state and the low-temperature
thermodynamics of the delta chain with F and AF interactions at
the transition point between the ferromagnetic and the
ferrimagnetic ground states. The most spectacular feature of this
frustrated quantum many-body system is the existence of a
macroscopically degenerate set of ground states leading to a large
residual entropy per spin of $s_0=\frac{1}{2}\ln 2$. Remarkably,
for these ground states explicit exact expressions can be found.
Among the exact ground states in the spin sector $S_{tot}=S_{\max
}-k$ there are states consisting of $k$ independent
(non-overlapping) magnons each of which is localized between two
neighboring apical sites. The same class of localized ground
states exist for the sawtooth model (\ref{q}) with both AF
interactions at the saturation field \cite{Zhitomir,derzhko,derz}.
However, such states do not exhaust all ground states in the
considered model. In addition to them, there are exact ground
states of another type consisting of products of overlapping
localized magnons. Since such states do not exist for the sawtooth
chain with both AF interactions,  in this respect the considered
model with F and AF interactions differs from the AF model. We
have checked our analytical predictions for the degeneracy of the
ground states in the sectors $S_{tot}=S_{\max }-k$ by comparing
them with numerical data for finite chains. The ground-state
degeneracy grows exponentially with the system size $N$ and leads
to above mentioned finite entropy per site at $T=0$. A
characteristic property of the excitation spectrum of the
$k$-magnon states is the sharp decrease of the gap between the
ground states and the excited ones when $k$ grows. As a result
both the highly degenerate ground-state manifold as well as the
low-lying excited states contribute substantially to the partition
function, especially at small $T$. That is confirmed by the
comparison of the data for the magnetization $M$ and the
susceptibility $\chi$ obtained by ED of finite chains with those
given by the contribution of the only degenerate ground states.
The subtle interplay of ground states and excited states leads to
unconventional low-temperature properties of the model. We have
shown that the magnetization $M$ at small $h$ and $T$ is a
function of the universal variable $h/T^{\alpha }$ with an index
$\alpha =1.09\pm 0.01$. This value of $\alpha $ agrees with the
critical index for the susceptibility. Furthermore, we have
analyzed the behavior of $\chi $ for finite chains. We have found
that this behavior can be described by one universal finite-size
scaling function. The entropy and the specific heat have also been
calculated by ED for finite chains. The entropy per site is finite
at $T=0$ and increases approximately with a power-law dependence
at $T>0$. The temperature dependence of the specific heat has a
rather interesting form characterized by a broad maximum at
$T\simeq 0.7$ and two weak maxima at $T\leq 0.1$.

Similar as the model with both AF interactions there is an
enhanced magnetocaloric effect. While for AF model this enhanced
effect is observed when passing the saturation field, we find it
for the considered model when the applied magnetic field is
switched off, which is obviously more suitable for a possible
application.

In conclusion, we note that the structure of the ground state
formed by the localized magnons is realized not only in the
critical point of the spin-$1/2$ F-AF delta-chain but also in the
$s_1,s_2$ chain, where $s_1$ and $s_2$ are the spins on the apical
and the basal sites correspondingly. The critical point for this
model is $\alpha_c=s_1/2s_2$ and the ground state in this critical
point has the same degeneracy as for the $s=1/2$ chain.


\begin{thebibliography}{9}
\bibitem{remark} This is very advantageous from an experimental point of
view with the aim to study these localized magnon states, since to
study them at frequently high saturation fields exceeding 40~Tesla
using pulsed fields, only, is this way circumvented.
\end{thebibliography}
\end{document}